\documentclass[final,5p,times,twocolumn]{elsarticle}

\usepackage{amssymb}
\usepackage{amsmath}
\usepackage{amsfonts}
\usepackage{graphicx}
\usepackage{dcolumn}
\usepackage{bm}
\usepackage{color}
\usepackage{xfrac}
\usepackage{tikz}
\usepackage{float}
\usepackage{hyperref}

\hypersetup{
    colorlinks=true,
    linkcolor=blue,
    citecolor=blue,
    urlcolor=blue
}

\journal{Physics Letters B}

\begin{document}

\begin{frontmatter}

\title{Topological dark energy from spacetime foam: A challenge for $\Lambda$CDM}

\author[1]{Fotios K. Anagnostopoulos}
\ead{fotis-anagnostopoulos@hotmail.com}

\author[2,3]{Stylianos A. Tsilioukas}
\ead{tsilioukas@gmail.com}

\author[3,4,5]{Emmanuel N. Saridakis}
\ead{msaridak@noa.gr}

\address[1]{Department of Informatics and Telecommunications, University of Peloponnese, Karaiskaki 70, 22100, Tripoli, Greece}
\address[2]{Department of Physics, University of Thessaly, 35100 Lamia, Greece}
\address[3]{National Observatory of Athens, Lofos Nymfon, 11852 Athens, Greece}
\address[4]{CAS Key Laboratory for Researches in Galaxies and Cosmology, Department of Astronomy, University of Science and Technology of China, Hefei, Anhui 230026, P.R. China}
\address[5]{Departamento de Matem\'{a}ticas, Universidad Cat\'{o}lica 
del 
Norte, 
Avda.
Angamos 0610, Casilla 1280 Antofagasta, Chile}

\begin{abstract}
Using only the standard considerations of spacetime foam and the Euclidean Quantum Gravity techniques known long ago, we result to a model of Topological Dark Energy (TDE) that competes the standard $\Lambda$CDM paradigm with regard to data fitting efficiency, while providing a microphysical mechanism for Dark Energy. Specifically, it is known that at the foam level, topologically non-trivial solutions such as instantons appear. In the particular case of Einstein-Gauss-Bonnet gravity, we obtain an effective dynamical Dark Energy term proportional to the instanton density, and the latter can be easily calculated through standard techniques. Hence, we can immediately extract the differential equation that determines the evolution of the topologically induced effective dark energy density. Significantly, this TDE scenario allows for changing sign of Dark Energy during the cosmic evolution and also exhibits Dark Energy interaction with Dark Matter. We confront the TDE scenario, in both flat and non-flat cases, with Pantheon+ Supernovae Type Ia (SNIa), Baryonic Acoustic Oscillations (BAO), and Cosmic Chronometers (CC) datasets. By applying standard model selection methods, we find the TDE scenario to be statistically equivalent with  $\Lambda$CDM. Finally, we show that the TDE scenario passes constraints from Big Bang Nucleosynthesis (BBN) and thus does not spoil the thermal history of the Universe.
\end{abstract}

\begin{keyword}
Dark Energy \sep Spacetime Foam \sep Gauss-Bonnet Gravity \sep Cosmology \sep Instantons
\end{keyword}

\end{frontmatter}


\section{Introduction}
Although the concordance $\Lambda$CDM paradigm is very 
succesful in describing 
the post-inflationary Universe at both background and perturbative levels, it 
exhibits theoretical and observational issues, such as the 
cosmological constant problem and also the $H_0$ and $\sigma_8$ tensions  \cite{Abdalla:2022yfr}. Hence, in the literature one can 
find a huge number of alternative, extended and modified theories and scenarios 
beyond $\Lambda$CDM model and/or general relativity  aiming to cure or alleviate 
the above disadvantages \cite{CANTATA:2021ktz,CosmoVerse:2025txj}.

On the other hand,   the attempt  to study the quantum behavior of gravity 
at the Planck scale has led to the concept of spacetime 
foam \cite{Wheeler:1955zz,Wheeler:1964}, where quantum fluctuations induce 
transient topological features  
\cite{Rovelli:2004tv,Green:1987sp,Hawking:1978jn,Coleman:1988tj}. Specifically, 
in the framework of Euclidean Quantum Gravity (EQG) at the 
foam level one has in general the appearance of solutions such as instatons  \cite{Hawking:1978pog}, which exhibit different topology from the 
background  \cite{Gibbons:2011dh}.

Although the above   features of spacetime foam are general, one can determine 
their exact behavior by choosing a specific gravitational theory. 
In a recent work we showed that the consideration of a   Gauss-Bonnet   
term  yields an effective cosmological constant 
of topological origin which is proportional to the density of gravitational 
instantons \cite{Tsilioukas:2023tdw}. In this Letter we intend to formulate the cosmological evolution of the instanton density and confront the 
resulting cosmological phenomenology with observations. In particular, by 
implementing standard Quantum Field Theory (QFT) techniques for the 
nucleation rate of instantons, that are known long ago, we derive the 
differential equation for the corresponding  Dark Energy (DE) density parameter 
$\Omega_{\Lambda_{\text{eff}}}$. The aforementioned equation, can be considered the second part of the theoretical framework for the Topological DE scenario.
It is worth noting that, in the context of Bayesian likelihood analysis, we find that 
the scenario at hand is {equivalent with $\Lambda$CDM regarding the  
CC/Pantheon+/BAOs and Pantheon+/BAOs datasets.

\section{Topological Dark Energy}
Let us present the details of the scenario of topological Dark Energy 
\cite{Tsilioukas:2023tdw}. According to the standard interpretation of Euclidean 
Quantum Gravity (EQG), topologically non-trivial 
gravitational instantons appear at the foam level, causing a change in the 
topology of spacetime, which can be evaluated by the connected sums formula $    
\delta\chi(M) = \chi(M_{\textrm{inst}}) - 2$
\cite{Gibbons:2011dh}, where the topological  index $\chi$ is the Euler 
characteristic, $M$ is the manifold of spacetime and $M_{\textrm{inst}}$ is the instanton 
manifold.  Each instanton 
species (i)  produces a distinct change, $\delta \chi_{i}$, which can be positive 
or negative \cite{Gibbons:1978ac,Gibbons:1979xm} (e.g a Nariai 
instanton leads to $\delta \chi = 2$, Euclidean wormholes yields $\delta 
\chi = -2$, etc \cite{Gibbons:2011dh,Tsilioukas:2023tdw}).

The above   features of spacetime foam are general, nevertheless in this 
work we are interested in examining their detailed behavior assuming a specific 
modification of gravity. In particular, we consider   the Einstein-Hilbert 
action plus the Gauss-Bonnet (GB) 
contribution in Euclidean signature, namely 
\begin{equation}
\label{actionEGB}
    I = I_{EH} + I_{GB} = \frac{1}{16\pi G}\left(\int d^{4}x \sqrt{g}  R + 
\alpha \int d^{4}x \sqrt{g}\;\mathcal{G}\right),
\end{equation}
where $\mathcal{G} = R^2 - 4 
R_{\mu\nu} R^{\mu\nu} + R_{\mu\nu\rho\sigma} R^{\mu\nu\rho\sigma}$ is the 
Gauss-Bonnet term, with   
$R_{\mu\nu\rho\sigma}$, $R^{\mu\nu}$, and $R$ the Riemann tensor the Ricci 
tensor and the  Ricci scalar respectively, and $\alpha$ is the corresponding GB 
coupling. 
If 
one splits the metric to the background metric and a quantum 
fluctuation,  namely $g_{\mu\nu}=\tilde{g}_{\mu\nu} + h_{\mu\nu}$, assumes that 
the quantum fluctuations encapsulate the EQG procedure of topology change and
thus causing a change in the Euler characteristic of spacetime $\delta h 
\rightarrow \delta \chi$, and moreover employs the semiclassical approximation, 
 then the Einstein equation for the background is obtained \cite{Tsilioukas:2023tdw}
\begin{equation}
\label{eq:einstein-field-1}    
\tilde{R}_{\mu\nu}-\frac{1}{2}\tilde{g}_{\mu\nu}\tilde{R}+\Lambda_{\textrm{eff}}
\tilde{g}_{\mu\nu}=8\pi G T_{\mu\nu}.
\end{equation}
In the above equations we have the appearance of an effective cosmological 
constant, $\Lambda_{\textrm{eff}}$, proportional 
to the rate of spacetime topology change per four volume, i.e.
\begin{equation}
\label{eq:effect_Lambda_chi}
    \Lambda_{\textrm{eff}} = - 16 \pi^2 \alpha \frac{\partial \chi}{\partial V}.
\end{equation}
Since $\delta\chi_{i}$ corresponds  to the appearance of an instanton, $\partial 
\chi/ \partial V$ can be estimated as the weighted sum density of instantons per 
four volume $n_i =  N_{\textrm{inst}}/V$ with weight $\delta\chi_{i}$, namely
\begin{equation}
\label{eq: Lambda sum}
    \Lambda_{\text{eff}} = - 16 \pi^{2} \alpha \sum_{i} \delta\chi_{i} n_{i}.
\end{equation} 
Observe that, from eq. \eqref{eq: Lambda sum}, the TDE scenario allows for 
changing sign of $\Lambda_{\text{eff}}$ during the cosmic history, as different instanton 
species (with different $\delta \chi_i$ sign) can coexist and/or suppress each 
other. 

In summary,  by combining the EQG spacetime topology change at the 
foam level, with the GB topological action term, without any other assumption 
one   obtains  the Einstein 
field equations  with an effective $\Lambda_{\text{eff}}$ term that is 
proportional to the density of the topology alternating instantons. One can now 
see the significance of the GB term in action (\ref{actionEGB}), since if it is 
absent then no $\Lambda_{\text{eff}}$ appears.

In order to apply the above model at a cosmological level, we consider a 
homogeneous and isotropic 
Friedmann-Robertson-Walker (FRW) geometry with metric
$
\mathrm{d}s^2 = -\mathrm{d}t^2 + a^2(t)  \left[ \mathrm{d}r^2(1 - k r^2)^{-1} 
+ r^2 \mathrm{d}\Omega^2 \right]$,
where \(a(t)\) is the scale factor, and \(k = 0, +1, -1\) corresponds to flat, 
closed, and open spatial geometry, respectively.
In this case, the field equations (\ref{eq:einstein-field-1}) give rise to the 
two Friedmann equations  
\begin{align}
\label{eq:friedman_1}
    H^{2}&=\frac{8 \pi G}{3}(\rho_{m}+\rho_{r}+\rho_{\text{DE}}) - \frac{k}{a^2} \\
    \label{eq:friedman_2}
    3H^{2}&+2\dot{H}+\frac{2 k}{a^2}=-8 \pi G (p_{r}+p_{\text{DE}}),
\end{align}
with $H=\dot{a}/a$ the Hubble function and where the energy density of the 
topological dynamical Dark Energy (DE) sector is
$\rho_{\text{DE}}=\Lambda_{\text{eff}}/8 \pi G$, and 
we have included the energy density and pressure of matter and 
radiation sectors.

\begin{figure*}
\includegraphics[scale=0.38]{ 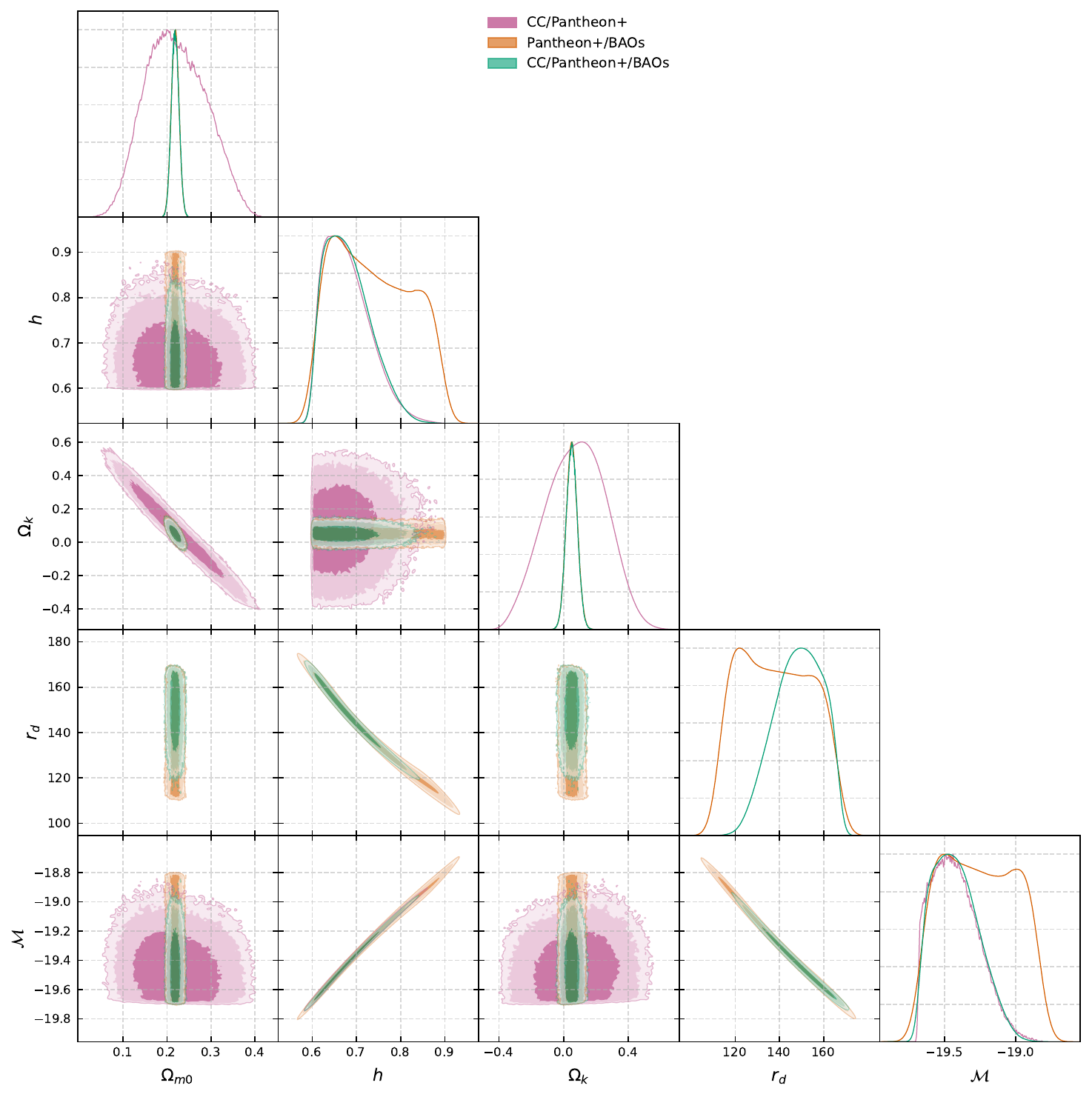}  
\includegraphics[scale=0.47]{ 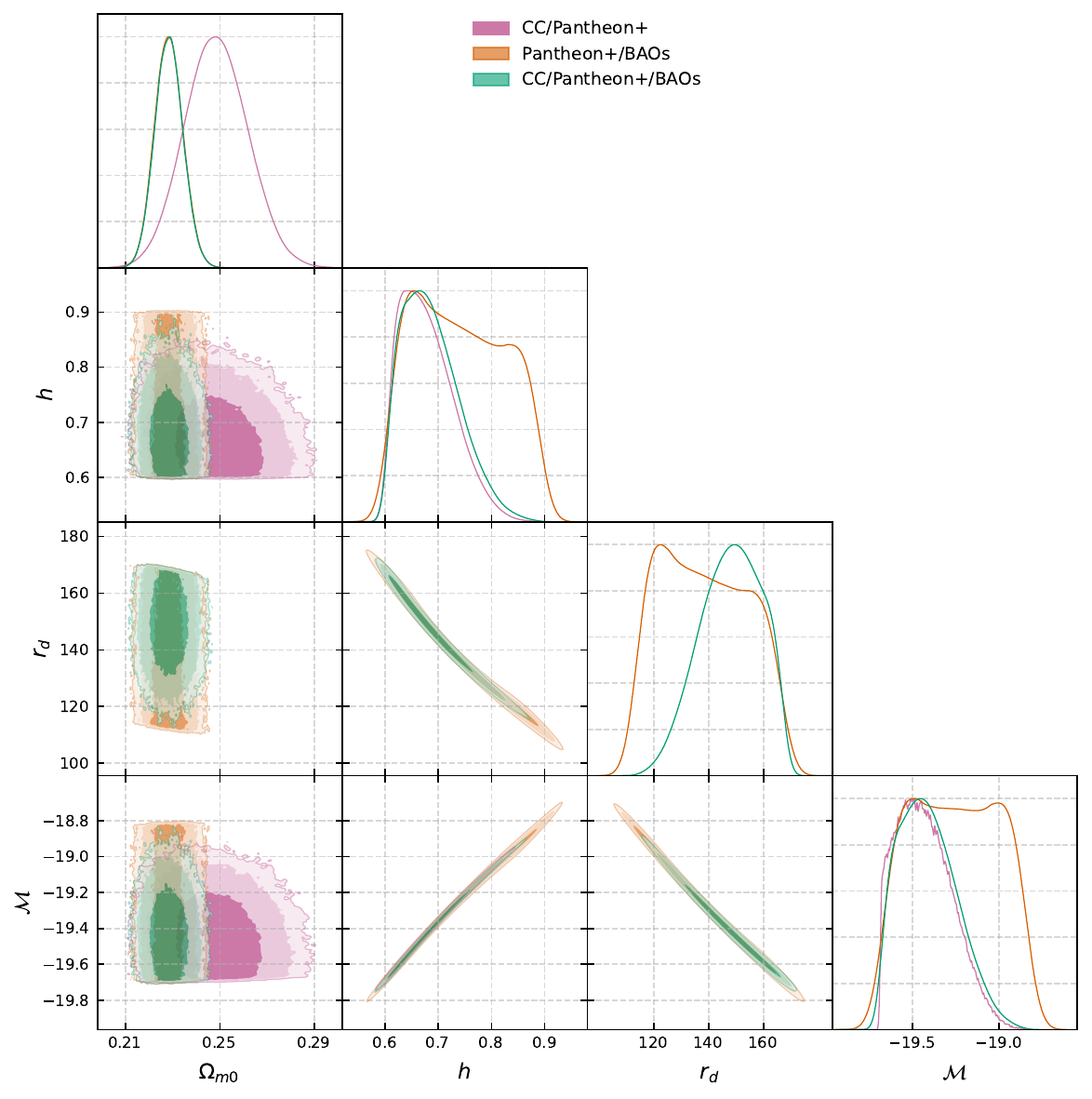} 
\caption[adf]{{\it{ Posterior distributions for parameter pairs, for all 
dataset 
compilations. The 
iso-surfaces correspond to 1$\sigma$-2$\sigma$-3$\sigma$ areas. Left graph: 
non-flat TDE model. Right graph: flat TDE model. } }}
\label{fig:non-flat-TDE-triangle} 
\end{figure*}

\begin{table*}[ht]
   \centering
    \caption{Parameter estimation results for flat and non-flat versions of the 
TDE scenario. Results for the concordance model are also included in order to 
allow for direct comparison.}
    \label{tab:parameters}
    \begin{tabular}{cccccccc}
        \hline \hline
        Model & $\Omega_{m0}$ & $\Omega_{k0}$ & $h$ & $r_d$ & $\mathcal{M}$ & 
$\chi_{\text{min}}^{2}$ & $\chi_{\text{min}}/dof$ \\
        \hline
        \multicolumn{8}{c}{CC/Pantheon+} \\
        \hline
        flat TDE & $0.248 \pm 0.013 $ & $-$ &$0.671_{-0.046}^{+0.059}$ & $-$ & 
$-19.44_{-0.15}^{+0.18}$ & $1393.40$ & $0.87$ \\
        non-flat TDE & $0.216_{-0.067}^{+0.077}$ & $0.083_{-0.197}^{+0.184}$ & 
$0.672_{-0.046}^{+0.060}$ & $-$ & $-19.44_{-0.15}^{+0.19}  $ & $ 1393.24$ & $0.87$ \\
        flat $\Lambda$CDM & $0.331_{-0.017}^{+0.018}   $ & $-$ 
&$ 0.671_{-0.046}^{+0.059} $ & $ - $ & $-19.44_{-0.15}^{+0.18}$ & $1393.40$ & $0.87$ \\
 non-flat $\Lambda$CDM & $ 0.304_{-0.054}^{+0.053} $ & $0.068_{-0.127}^{+0.131}$ 
&$0.669_{-0.045}^{+0.058}$ & $- $ & $-19.45_{-0.15}^{+0.18} $ & $1393.11$ & $0.87$ \\
        \hline
               \multicolumn{8}{c}{{Pantheon+/BAOs}} \\
        \hline
        flat TDE & $0.228 \pm 0.006 $ & $-$ &$0.730_{-0.091}^{+0.111} $ & $138_{-18}^{+19}$ & $-19.27_{-0.29}^{+0.31}$ & $1400.32$ & $0.88$ \\
        non-flat TDE & $0.219 \pm 0.009 $ &  $0.051 \pm 0.034$ & $0.7287_{-0.090}^{+0.111}$ & $ 138_{-18}^{+19}$ & $ -19.27_{-0.29}^{+0.31}$ & $1398.06 $ & $0.88$ \\
         flat $\Lambda$CDM & $ 0.304 \pm 0.008$ & $-$ & 
$0.732_{-0.093}^{+0.108}   $ & $138_{-17}^{+20} $ & $ -19.3_{-0.3}^{+0.3} $ & 
$1400.35 $ & $0.88 $ \\
 non-flat $\Lambda$CDM & $0.293_{-0.011}^{+0.012}  $ & $0.049_{-0.036}^{+0.037} $ & 
$0.732_{-0.094}^{+0.110}   $ & $137_{-18}^{+20}  $ & $-19.3_{-0.4}^{+0.3}$ & 
$1398.61 $ & $0.88$ \\
         \hline

        \multicolumn{8}{c}{CC/Pantheon+/BAOs} \\
        \hline
        
        flat TDE & $0.228 \pm 0.006 $ & $-$ & 
$0.680_{-0.050}^{+0.061} 
$ &  $148_{-12}^{+11}  $ & $-19.43_{-0.16}^{+0.19}$ &  $1406.72$ &  $0.87 $\\
        non-flat TDE  & $0.219 \pm 0.008 $ &  $0.050_{-0.033}^{+0.034} $ & 
$0.674_{-0.048}^{+0.060}  $ & $149_{-12}^{+11}  $ & $-19.44_{-0.16}^{+0.18}  $ & $1404.46$ & $0.87 $ 
\\
       flat $\Lambda$CDM &   $0.304 \pm 0.008 $   & $-$ &  
$0.679_{-0.050}^{+0.062}$   &  $148_{-12}^{+11}$ &  $-19.43_{-0.16}^{+0.19}$ &  
$1406.76$ &  $0.87$ \\
  non-flat $\Lambda$CDM &  $0.293 \pm 0.011$  & $0.047 \pm 0.036  $ &  
$0.674_{-0.048}^{+0.060}$   &  $148_{-12}^{+11}$ &  $-19.43 _{-0.16}^{+0.18}$ &  
$1405.03$ &  $0.87$ \\
        \hline
        \hline
    \end{tabular}
\end{table*}

Since the topological   DE density is   proportional to the density of 
instantons, which can in principle vary with time, in order to proceed we need 
to calculate the latter in a cosmological background. The appropriate 
theoretical framework for describing the nucleation process of gravitational 
instantons originates from bubble nucleation theory \cite{Kobzarev:1974cp}, 
developed to describe the tunneling process from false to true vacuum 
\cite{Coleman:1977py,Coleman:1980aw}.  Specifically, the probability per unit 
volume per unit time for an instanton to occur is given by  
$    \Gamma = A \ \textrm{exp}(-\Delta I)$,
where  $\Delta I $ is  the difference in the Euclidean action between the 
instanton (tunneling) configuration and the surrounding background 
configuration 
\cite{Coleman:1977py,Coleman:1980aw,Gibbons:1976ue,Hawking:1981fz,
Gibbons:1992rh}. Moreover, the quantity \(A\) has been calculated at Ref. 
\cite{Coleman:1977py,Coleman:1980aw}. Since a tunneling process corresponds to 
an instanton, the rate $\Gamma$  can be interpreted as the density of 
instantons per four volume, namely $n \equiv \Gamma$. Hence, if we allow for  $i$ different species of instantons 
we  acquire
\begin{equation}
\label{eq: n eq Gamma}
    n_{i} \equiv \Gamma_{i} = A_{i} e^{-\Delta I_{i}}.
\end{equation}
Let us calculate  $\Delta I$ for a given instaton specie. In the case of 
Einstein-Gauss-Bonnet theory (\ref{actionEGB}),  the action difference $\Delta 
I$ for each type of instanton consists of a topological and a geometrical 
contribution $\Delta I = \Delta I_{\text{GB}} + \Delta I_{\text{EH}}$.  
For the first term, applying the Chern-Gauss-Bonnet theorem 
\cite{Chern1945OnTC} 
$
    \chi(M) = 1/(32 \pi^{2})\int d^{4}x \sqrt{g} \;\mathcal{G}$,
we find that the 
topological contribution for each instanton species $(i)$ is given by 
$
    \Delta I_{\text{GB}}(i) = 2\pi \alpha \chi_{i}G^{-1}$.
For the second term, namely $\Delta I_{\text{EH}}$, we start by mentioning that 
since gravitational instantons are vacuum 
solutions, the corresponding Ricci scalar is $R_{\textrm{inst}}=4 \Lambda$ 
\cite{Parikh:2009js}, where in our case $\Lambda=\Lambda_{\text{eff}}$.
On the other hand, the FRW background metric has a Ricci 
scalar $R_{\textrm{back}}=6(2 H^{2}+\dot{H}+k/a^2)$. In our approach, we 
consider instantons nucleated at the foam level within a cosmological 
background, therefore we assume the same four-volume of integration for the 
background and the instantons defined from the FRW metric of Lorentzian 
signature, thus the geometrical part $\Delta I_{EH} = (16\pi G)^{-1} \int 
d^{4}x \sqrt{g} (R_{inst} - R_{back})$ is the same for all instantons.
Hence, assembling both terms we find that 
{\small{
\begin{equation}
 \Delta I =  \frac{2\pi \alpha \chi_{i}}{G}+ \frac{1}{16\pi G} \int 
d^{4}x \sqrt{g} \left[4 \Lambda_{\text{eff}}-6\left(2 
H^{2}+\dot{H}+\frac{k}{a^2}\right)\right],
\end{equation}}}
\noindent
with $\Lambda_{\text{eff}}$ given in (\ref{eq: Lambda sum}). Hence, inserting this 
expression into (\ref{eq: n eq Gamma}) gives the instantons density $ 
n_{i}$. As a final step we consider that the  spatial part of the Universe four volume is a Hubble sphere of radius $r=1/H$.
Substituting the above 
obtained  $n_{i}$ into  Eq.~\eqref{eq: Lambda sum}, and differentiating,  we 
obtain  a differential equation for  $\Lambda_{\text{eff}}$, namely
\begin{equation}
\label{eqaux1}
    \frac{d\Lambda_{\text{eff}}}{dt} =\frac{1}{12 G} \frac{a^{3}}{H^{3}} \left( 
12 H^{2} + 6\frac{dH}{dt} +6\frac{k}{a^2} - 4 
\Lambda_{\text{eff}}\right)\Lambda_{\text{eff}},
\end{equation}
where we have assumed that $A_{i}$ does not depend of the cosmic time.
As usual, in the case of dynamical dark energy, a convenient expression for the 
dimensionless Hubble rate ($E(z) \equiv H(z)/(100 \cdot h)$) is 
\begin{equation}
\label{eq:E-rate}
    E(z) = \left[\frac{ 
\Omega_{r0}(1+z)^{4}+\Omega_{m0}(1+z)^{3}+\Omega_{k0}(1+z)^{2}}{1-\Omega_{
\Lambda_{\text{eff}}}(z)}\right]^{1/2},
\end{equation}
where we have introduced the redshift through  $dt = -dz(1+z)^{-1} H^{-1}$.
Hence, (\ref{eqaux1}) finally yields
\begin{align}
\label{eq:ode_model_z}
    &\frac{d\Omega_{\Lambda_{\text{eff}}}(z)}{dz} = (\Omega_{\Lambda_{\text{eff}}}(z)-1)\Omega_{\Lambda_{\text{eff}}}(z) \bigg[4 G H_{0}^{2} (z+1)^{5} f_{1}(z) \nonumber
    \\
    &\cdot f_{2}(z) + (1-\Omega_{\Lambda_{\text{eff}}}(z))\cdot\big(  \Omega_{m0} (z+1) -4 (f_{1}(z)-\Omega_{k0})  \nonumber
    \\
    &\cdot\Omega_{\Lambda_{\text{eff}}}(z)  \big)\bigg]\cdot \bigg[ (z+1) f_{1}(z) \big(4 G H_{0}^{2} (z+1)^{5} f_{1}(z)\nonumber
    \\
    & - (1 - \Omega_{\Lambda_{\text{eff}}}(z) )\Omega_{\Lambda_{\text{eff}}}(z) 
\big)\bigg]^{-1},
\end{align}
where for convenience we have defined
\begin{align}
    f_{1}(z) \equiv\ & \Omega_{k0} + (z + 1) \left[\Omega_{m0} + \Omega_{r0} (z+1)\right] \\
    f_{2}(z) \equiv\ & 2 \Omega_{k0} + (z+1) \left[3 \Omega_{m0} + 4 \Omega_{r0} (z+1)\right].    
\end{align}
Equation (\ref{eq:ode_model_z}) is the differential equation that determines 
the evolution of topological dark energy (TDE), and can be 
solved numerically.  At z = 0, the normalization condition $E(z=0) = 1$ imposes $\Omega_{\Lambda_{\textrm{eff}}}(z=0) = 1 - \Omega_{m0} - \Omega_{r0}-\Omega_{k0}$, which serves as an initial condition. Note that the initial condition determines the value of $\Omega_{\Lambda_{\textrm{eff}}}$, thus the instanton species mix at eq. \eqref{eq: Lambda sum}.
Finally, considering 
the continuity equation for the DE  species, i.e. 
$\dot{\rho}_{\text{DE}}+3H(1+    w_{\text{DE}})\rho_{\text{DE}}=0$,  we extract 
the expression for the DE equation-of-state parameter as
\begin{equation}
\label{eq: top_DE_wDE}
    w_{\text{DE}}(z)=-1+\frac{(1+z)}{3} 
\left[\frac{d\ln\Omega_{\Lambda_{\text{eff}}}(z)}{dz}+2\frac{d\ln 
H(z)}{dz}\right ] .
\end{equation}
The above conclude the development of the TDE model at the background level. Regarding the linear cosmological perturbations, a complete analysis within the TDE framework has not yet been derived for either flat or non-flat geometries. This absence is not a mere technical oversight but reflects deeper theoretical challenges that must be addressed to ensure a consistent model. Specifically, the framework requires a formal treatment of the backreaction of metric fluctuations onto the instanton nucleation rate $\eqref{eq: n eq Gamma}$ and a rigorous definition of the four-volume integration under perturbed spatial boundaries.
 
\section{Observational confrontation} 
We can now proceed to the investigation of the observational consequences, at the background level, of 
the proposed topological DE scenario, examining both the ``flat  
TDE'', i.e. by setting $\Omega_{k0} = 0$, as well as the ``non-flat TDE'', where 
$\Omega_{k0} \neq 0$. We use data from Type Ia Supernova (SNIa) observations, specifically a Pantheon+ subsample \cite{Scolnic2022PantheonPlus,scolnic2018complete} with $z > 0.01$. This cut allows us to circumvent the effects of model-dependent peculiar velocity corrections applied to the Pantheon+ data \cite{lane2025cosmological}. This is necessary because, in the very near-field universe, local cosmic structure can dominate the expansion signal, making such corrections highly influential. Also, we use
direct measurements of the Hubble function, namely Cosmic 
Chronometers (CC) data (see \cite{anagnostopoulos2024observational} and 
references therein) and data from Baryonic Accoustic Oscillations (BAOs). Specifically, we used the recent DESI DR2 BAO measurements \cite{abdul2025desiI, abdul2025desiII}, which provide a dense sampling of the cosmic expansion history for $0.1 < z < 4.2$. To prevent double-counting and systematic inconsistencies between survey pipelines, we rely exclusively on the DESI DR2 data rather than combining it with previous BAO datasets. In
order to obtain the posterior distributions of the model parameters we 
use an affine-invariant Markov Chain Monte Carlo (MCMC) 
sampler as implemented within the open-source Python package emcee 
\cite{foreman2013emcee,foreman2019emcee}, involving 1000 ``walkers'' (chains) and 
2500 ``states'' (steps), and regarding the convergence of the MCMC algorithm, 
we use the traditional Gelman-Rubin criterion and also the auto-correlation time 
analysis. Finally, in order to compare the statistical efficiency of the 
scenario at hand comparing to $\Lambda$CDM paradigm, we employ four widely 
recognized criteria: the Akaike Information Criterion (AIC), the Bayesian 
Information Criterion (BIC), the Deviance Information 
Criterion (DevIC)   \cite{liddle2007information} and an estimation for the Bayesian Evidence. To compute the latter from our MCMC chains, we utilized the MCEvidence package, which estimates the posterior volume using a $k$-Nearest Neighbor algorithm in the parameter space \cite{heavens2017marginal}. The subsequent model assessment can be done via the standard  
Jeffreys scale \cite{Kass:1995loi}. 
   
We depict 2D slices of the posterior distributions of the free parameters for the  non-flat and flat TDE models in Fig. \ref{fig:non-flat-TDE-triangle}, for 
the case of CC/Pantheon+, Pantheon+/BAOs and 
CC/Pantheon+/BAOs datasets.  Additionally, 
in  Table \ref{tab:parameters}  we summarize the results   compared against the 
standard flat $\Lambda$CDM ($\Omega_{k0}=0$) and the non flat $\Lambda$CDM ($\Omega_{k0} \neq 0$) cosmological models. 
In the case of the combined analysis of all    datasets  
(CC/Pantheon+/BAOs)
we find that both
the flat and non-flat TDE model favors a relatively low value of  
$\Omega_{m0}$, significantly smaller than the $\Lambda$CDM value of 
$\Omega_{m0} = 0.304 \pm 0.008 $.
Moreover, the non-flat TDE model accommodates a slightly higher curvature value $\Omega_{k0}$ 
with respect to the non-flat $\Lambda$CDM. However, both curvature energy-density values remain compatible 
at the $1\sigma$ level. The hyperparameter $\mathcal{M}$ remains nearly constant across different model combinations. Interestingly, the dimensionless Hubble constant, $h$, for the case of Pantheon+ and BAOs data is closer to the local measurement, albeit compatible with Planck's mission \cite{Aghanim:2018eyx} $h = 0.6736 \pm 0.0054$. Similarly, the sound horizon $r_d$ remains stable across models and is consistent with the value reported by the Planck mission \cite{Aghanim:2018eyx}, i.e. $r_d = 147.09 \pm 0.26 \text{ Mpc}$ , although with much larger uncertainties. We must note that in the fitting analysis with only the Pantheon+/BAOs datasets, the calibrators of the direct and inverse distance ladders are fully unconstrained by construction. Consequently, the large uncertainties observed for $h$, $r_d$, and $M$ in this specific analysis are artificial. They do not represent true physical bounds but rather reflect the prior boundaries and the finite length of the Monte Carlo Markov chains. Regarding the interpretation of the reduced 
$\Omega_{m0}$ in comparison with $\Lambda$CDM, it is of interest to plot the DE 
equation of state parameter in Fig. \ref{fig:wdeII}. The TDE scenario predicts 
that the DE equation of state evolves from 
$w_{DE} \approx 0$ at $z \sim 10$, mimicking pressureless Dark Matter (DM), to $w_{DE} = -0.89$ today. This behavior corresponds to an effective conversion between dark energy and dark matter. Such DE-DM mixing suggests that TDE models may alleviate both the $H_0$ and $\sigma_8$ tensions, similarly to phenomenological interacting DE/DM models \cite{DiValentino:2020vvd}. Furthermore, regarding the growth of structure, the clustering of dark energy actively acts as a repulsive agent. This reduces the overall clustering efficiency for a given matter density and subsequently breaks the classical anticorrelated degeneracy between $\Omega_{m0}$ and $\sigma_8$ typically observed in standard cosmological probes \cite{Batista2022}. In contrast to these phenomenological interacting scenarios, the TDE model originates from first principles, providing a concrete microphysical mechanism for the effective cosmological constant. The absence of a derived perturbation scheme currently precludes the integration of the TDE model into standard Boltzmann solvers, such as CLASS or CAMB. Developing such modifications requires a self-consistent treatment of these fluid transitions to accurately evolve the CMB power spectra \cite{class_issue425_comment2025}. Consequently, a full MCMC assessment against CMB data to assess the aforementioned possibility regarding $H_0$ and $\sigma_8$ tensions remains beyond the scope of this work.

In Tab. \ref{tab:Results2} we apply a number of information criteria (\cite{anagnostopoulos2024observational} and references therein), 
and we calculate the corresponding difference $\Delta$IC $\equiv 
 \text{IC} - \text{IC}_{\text{min}}$. For the case of the Bayesian evidence, $\Delta\ln\mathcal{Z} \equiv \ln\mathcal{Z}_{\max} - \ln\mathcal{Z}$. In all cases, $\Delta IC = 0$ corresponds to the best model in the list. In summary, analyzing the information criteria and the Bayesian evidence yields two primary conclusions. First, the $\Lambda$CDM and TDE models perform extremely similarly and cannot be distinguished from a statistical point of view, regardless of the information criterion used. Second, the only notable statistical differences emerge between the flat and non-flat geometries in both the $\Lambda$CDM and TDE frameworks. Ultimately, in the joint analysis (CC/Pantheon+/BAOs), the flat TDE model ($\Delta\ln\mathcal{Z} = 0.06$) and the standard flat $\Lambda$CDM model ($\Delta\ln\mathcal{Z} = 0.00$) perform almost identically. Thus, we conclude that the flat TDE model is an excellent alternative to the standard cosmological model, in the light of the datasets considered here.

\begin{table*}[ht]
   \centering
\tabcolsep 2.0pt
\vspace{2mm}
\begin{tabular}{ccccccccc} \hline \hline
Model & AIC & $\Delta$AIC & BIC &$\Delta$BIC & DIC& $\Delta$DIC & $\ln\mathcal{Z}$ & $\Delta\ln\mathcal{Z}$
 \vspace{0.02cm}\\ 
 \hline
\hline 
 \multicolumn{9}{c} {CC/Pantheon+}\\ 
  TDE &  $1399.41$ & $0.00$ & $1415.54$ & $0.00$ & $1396.10$ & $0.02$ & $1.06$ & $0.95$ \\ 
   non flat TDE &  $1403.27$ & $3.86$ & $1430.14$ & $14.60$ & $1396.94$ & $0.86$ & $2.01$ & $0.00$ \\
$\Lambda$CDM & $1399.42$ & $0.01$ & $1415.55$ & $0.01$ & $1396.08$ & $0.00$ & $0.97$ & $1.04$\\ 
 non flat $\Lambda$CDM  &  $1401.14$ & $1.73$ & $1422.63$ & $7.09$ & $1397.98$ & $1.90$ & $1.17$ & $0.84$ \\
 \hline

   \multicolumn{9}{c}{ {Pantheon+/BAOs}}\\
 \vspace{0.05cm}
 TDE &  $1408.34$ & $0.24$ & $1429.81$ & $0.00$ & $1367.30$ & $1.43$ & $2.78$ & $0.00$ \\ 
   non flat TDE &  $1408.10$ & $0.00$ & $1434.93$ & $5.12$ & $1366.99$ & $1.12$ & $1.11$ & $1.67$ \\
$\Lambda$CDM & $1408.38$ & $0.28$ & $1429.85$ & $0.04$ & $1367.15$ & $1.28$ & $2.78$ & $0.00$\\
 non flat $\Lambda$CDM  &  $1408.65$ & $0.55$ & $1435.48$ & $5.67$ & $1365.87$ & $0.00$ & $1.83$ & $0.95$ \\
 \hline
  \multicolumn{9}{c}{ {CC/Pantheon+/BAOs}}\\
 \vspace{0.05cm}
TDE &  $1414.74$ & $0.24$ & $1436.27$ & $0.00$ & $1406.61$ & $0.16$  & $2.65$ & $0.06$ \\ 
   non flat TDE &  $1414.50$ & $0.00$ & $1441.41$ & $5.14$ & $1406.97 $ & $0.52$ & $1.40$ & $1.30$ \\
$\Lambda$CDM & $1414.78$ & $0.28$  & $1436.31$& $0.84$ & $1406.45 $ & $0.00$ & $2.70$ & $0.00$\\
 non flat $\Lambda$CDM  &  $1415.07$ & $0.57$ & $1441.97$ & $5.70$ & $1407.32$ & $0.87$ & $1.12$ & $1.58$ \\
 \hline\hline
\end{tabular}
\caption{The information criteria (AIC, BIC, and DevIC) and the logarithm of the Bayesian evidence ($\ln\mathcal{Z}$) for the examined cosmological models. The corresponding differences are defined as $\Delta\text{IC} \equiv \text{IC} - \text{IC}_{\text{min}}$ and $\Delta\ln\mathcal{Z} \equiv \ln\mathcal{Z}_{\max} - \ln\mathcal{Z}$, ensuring that a value of $0.00$ universally indicates the statistically preferred model in each column. \label{tab:Results2}}
\end{table*}

We close our analysis by a consistency check from Big Bang Nucleosynthesis (BBN). The latter  can be 
performed via an estimation of the percentage increment of the dilation rate of 
the universe at BBN era, i.e $z_{\textrm{BBN}} \sim 10^9$, using the best fit values for 
the free parameters \cite{sola2017first}.
Calculating the   ratio 
$
    C_r =  (H_{TDE} - H_{\Lambda CDM})^2/H_{\Lambda CDM} ^2,
$
we find   that the BBN constraints are satisfied for all TDE models 
considered here.

\begin{figure} 
\includegraphics[scale=0.4]{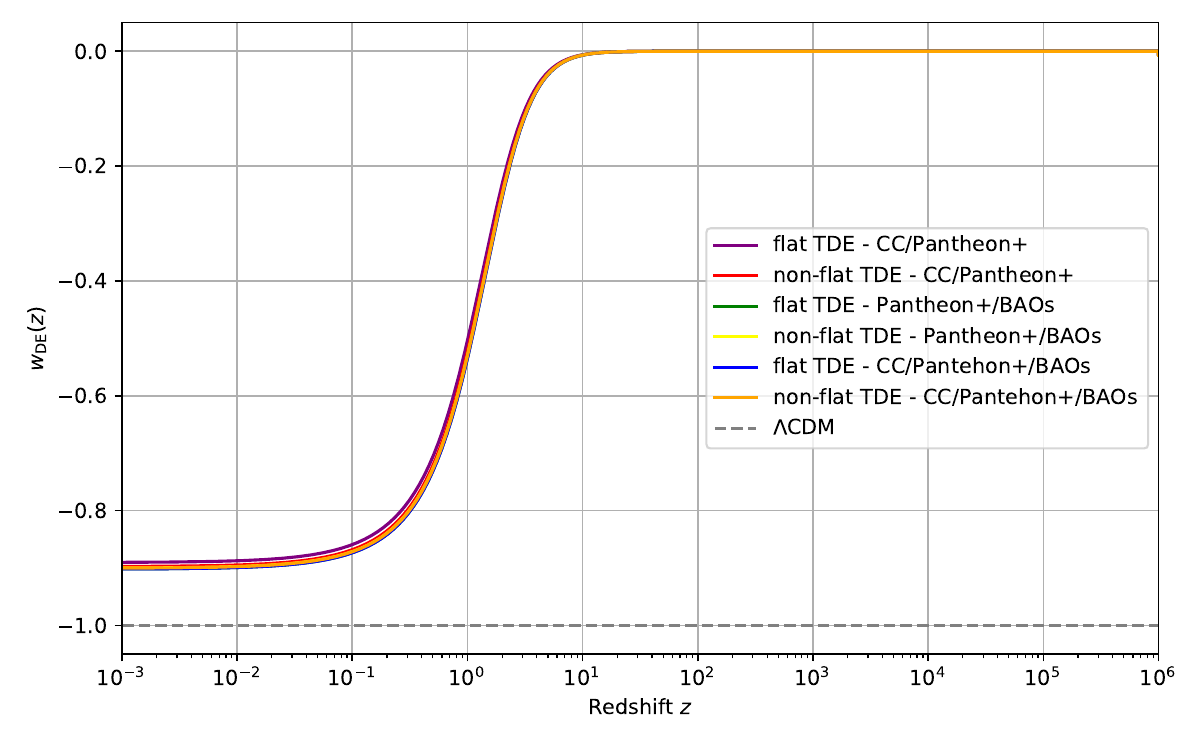}  
\caption[adf]{{\it{The evolution of the dark-energy equation-of-state parameter 
 $w_{DE}$, as a function of 
redshift from Eq.~\eqref{eq: top_DE_wDE}, for the flat and the non-flat TDE 
model, with parameter values from the best fit in the three datasets as they are 
presented in Table~\ref{tab:parameters} .}}}
\label{fig:wdeII} 
\end{figure}
 
\section{Conclusions}
This work presents a novel Topological Dark Energy (TDE) scenario, which involves a microphysical mechanism for DE production. When tested against current observations, the model demonstrates a statistical performance that rivals the standard $\Lambda$CDM paradigm across the most recent background cosmological datasets. The TDE model is based only on the standard considerations of spacetime foam and Euclidean Quantum Gravity techniques known long ago. Specifically, it is known that topologically non-trivial solutions, such as instantons, emerge at the level of spacetime foam in the aforementioned context. In the particular case of an enhanced gravitational action with the Einstein-Gauss-Bonnet term,
one obtains an effective dynamical DE term proportional to the 
instanton density, and the latter can be easily calculated through standard 
techniques. Hence, one can easily extract the differential equation that 
determines the evolution of the topological DE density parameter. Within the TDE scenario, the effective cosmological constant 
becomes dynamical, while interestingly enough, is allowed to change sign 
throughout cosmic history. The TDE scenario incorporates dark sector interactions, a feature connected with possible alleviation of the $H_0$ and $\sigma_8$ tensions, at least as seen in similar interacting models. 
We confronted the TDE scenario, in both flat and non-flat cases, with   
Pantheon+, BAO, and Cosmic Chronometers (CC) datasets and we extracted 
the constraints on the model parameters. Despite the inclusion of various information criteria, the TDE model remains statistically consistent with $\Lambda$CDM, showing no evidence of inferiority in its ability to describe the observed data.

Topological Dark Energy arises from basic first principles about the properties 
of spacetime foam. The fact that such a simple scenario can lead to a 
statistically equivalent cosmological behavior than the concordance $\Lambda$CDM  
paradigm, makes it a promising and viable alternative that deserves further 
investigation. To complete this picture, a full treatment of TDE perturbations towards utilization of CMB data is currently in progress.

\bibliographystyle{elsarticle-num}
\bibliography{PLB}

\end{document}